\documentclass{article}

\usepackage{arxiv}

\usepackage[utf8]{inputenc} % allow utf-8 input
\usepackage[T1]{fontenc}    % use 8-bit T1 fonts
\usepackage{hyperref}       % hyperlinks
\usepackage{url}            % simple URL typesetting
\usepackage{booktabs}       % professional-quality tables
\usepackage{amsfonts}       % blackboard math symbols
\usepackage{nicefrac}       % compact symbols for 1/2, etc.
\usepackage{microtype}      % microtypography
\usepackage{lipsum}		% Can be removed after putting your text content
\usepackage{float}
\usepackage{graphicx}

\title{An AI based talent acquisition and benchmarking for job}

\author{ Rudresh Mishra\\
	MSc. Data Science\\
	IMT Atlantique\\
	Brest, France \\
	\texttt{rudresh.mishra@imt-atlantique.net}
\And
	Ricardo Rodriguez \\
	MSc. Data Science\\
	IMT Atlantique\\
	Brest, France \\
	\texttt{ricardo-moises.rodriguez-oceda@imt-atlantique.net}
\And
	Valentin PORTILLO\\
	MSc. Data Science\\
	IMT Atlantique\\
	Brest, France \\
	\texttt{valentin.portillo-galvan@imt-atlantique.net}
	}
\hypersetup{
pdftitle={An AI based talent acquisition and benchmarking for job},
pdfsubject={Computer science, NLP, CV matching},
pdfauthor={Rudresh Mishra,Ricardo Rodriguez, Valentin PORTILLO},
pdfkeywords={AI, Resume, Human Resource Management, Recruitment, Graph Theory, Machine learning, Natural language.}
}

\begin{document}
\maketitle
\begin{abstract}
In a recruitment industry, selecting a best CV from a particular job post within a pile of thousand CV’s is quite challenging. Finding a perfect candidate for an organization who can be fit to work within organizational culture is a difficult task. In order to help the recruiters to fill these gaps we leverage the help of AI. We propose a methodology to solve these problems by matching the skill graph generated from CV and Job Post. In this report our approach is to perform the business understanding in order to justify why such problems arise and how we intend to solve these problems using natural language processing and machine learning techniques. We limit our project only to solve the problem in the domain of the computer science industry.
\end{abstract}


% keyword change to index

\section{Introduction}

Unlike traditional recruitment methods, such as employee referrals, CV screening, and face-to-face interviews, AI is able to find patterns unseen by a human eye. It could be used to find the right person for the perfect role faster and more efficiently than ever before. In order to rapidly improve talent management and take full advantage of the power and potential AI offers, we need to shift our focus from developing more ethical HR systems to developing more ethical AI. McKinsey's Global Institute model predicts that approximately 70 percent of companies will adopt some form of AI by 2030. When it comes to identifying talent or potential, most organizations still play it by ear. Recruiters spend just a few seconds looking at a resume before deciding who to “weed out” \cite{erecruit_cv_form}. Often when hiring is made it's very important to know the current strength of the organization and based on it if hiring is made a candidate is referred to be a good fit for an organization \cite{sel_soc_acc}. There's increasing evidence that AI could overcome this trade-off  by deploying more dynamic and personalized scoring algorithms that are sensitive as much accuracy as to fairness to an organization.

AI has power to provide deep hiring efficiencies, increase talent mobility and will ensure that the scores that come out of the hiring process are both maximally predictive of outcomes that matter to employers, free from all types of bias and provides the best fitting candidate as per organizational work environment. AI and ML have an immense potential to provide a unique solution in the domain varying from robotic automation to biochemical industry\cite{Micro_nanorobots} \cite{Ai_ml_fornano_tech}.

Recent interesting work where Real Time Heart Rate Measurement with Facial Video has been performed using face detection technique. This approach can also be implemented in the Hr tech industry in order to interview the candidate to know more about their behavior\cite{Singh2017ContactlessAH}.  

In this report, we present an analysis of the problem of recruitment, followed by a proposition of the model. In Section 2, we will cover a general understanding of the recruitment field, and we will cover the problem statement of this project and the Implementation.

According to several surveys, the recruiting field is one of the main concerns of many CEO’s \cite{hiring_approach, ceo_survey_talent}. In fact, according to the Society for Human Resource Management \cite{hiring_approach}, employers spend an enormous amount on hiring: an average of \$4,129 per job in the United States.

We observe that the recruiting process is not an easy task. It contains several stages. The average time to fill an open position is approximately 42 days \cite{avg_hire_cost} and even with this long process, most of the time recruiters are not sure that they choose the right candidate \cite{hiring_approach}.

To overcome this issue, many innovations in the recruiting field have arisen recently, such as video interviews analysis, accurate CV parsers, AI personality tests, AI candidate recommendation, among others. According to an analysis made by the Linkedin talent blog in 2018 \cite{g_recruiting_trends}, there exists four trends that will shape the future of recruiting:

\begin{list}{--}{}
\item  \textbf{Diversity}: Refers to the fact that changing demographics are diversifying communities, shrinking talent pools for companies that don’t adapt. This trend is relevant since diverse teams are more productive, more innovative, and more engaged also makes it hard to ignore.
\item \textbf{New interviewing tools}: These tools try to improve ineffective traditional ways of interviewing. New tools are concentrated on online soft skills assessments, job auditions, casual interviews, among others.

\item \textbf{Data} : Refers to data informing talent decisions, such as prediction of hiring outcomes, or smarter recruiting decisions based on data analysis.

\item \textbf{Artificial Intelligence}: It is focused on automated candidate searches and quickly finds prospects that match specific criteria. There are also technologies that help to screen candidates before even speaking to them. The development of chatbots can respond to candidate questions so recruiters don’t have to.

\end{list}

While Sourcing candidates is the process to contact as many candidates as possible, screening candidates refers to the problem of selecting a candidate based on its CV. It makes sense that the screening is one of the fields where innovation is needed, since Profile/CV matching is a multi dimensional task. Normal human eye is not enough to compare precisely many CV's in a multi dimensional way.

\section{Problem Identification and Objectives}
\subsection{Problem}

Recruitment can be a very demanding and tough process for a company and their recruiters. Many times, recruiters end up hiring a not so competent candidate which eventually renders all the efforts put through a recruitment process as waste \cite{algo_recruiter}. Having a perfect fit for a job position is as tough as finding that perfect fit and that entire process of finding one can be very demanding at times. It is very important for a recruiter to pick a candidate whose competent matches with current organization strength. In addition to this, there are a lot of difficulties which candidates face while searching for their dream job. Starting from finding a trusted platform to searching for job roles, to track their application and to receive feedback for application, the entire process has a lot of roadblocks which renders the entire recruitment process as very time consuming and as a frustrating one \cite{web_based_recruiting}. The root problem is because the profile matching field is multi dimensional and it is very difficult as an individual to cover all dimensions and to select the best candidate and to justify the reason for the selection and rejection of the candidate.

As a way out, the candidates or recruiter often make use of third parties to reach out for their desired job roles and positions. they have a team dedicated to a more manual approach to do the matching. This eventually results in a major chunk of their salary being lost to the third party facilitators and targeted problems cannot be solved.

Every organization works as per their unique values and strengths and it's very hard to generalize the common matching solution to a candidate which can fit best for all the organization. There are many solutions existing in the market to automate the matching such as creating the recommendation systems which are based on keyword matching which often results in poor recommendations. Also there are many AI related solutions which provide solutions to the problem, however if the candidate is recruited without considering the organization values and strength, it becomes hard for a candidate to survive and give best to an organization.

\subsection{Current market solution}

The table \ref{table:2} shows an AI driven talent platform which has been assisting the enterprise along with the features details.

\begin{table}[h!]
\caption{AI driven talent platform h}
\label{table:2}
  \centering
  \begin{tabular}{p{3cm} p{8cm}} 
    \hline\hline
    Metric & Description  \\
    \hline
    Text kernel &
    ML (DL) for document understanding, Web Mining external sources, Synonyms,Software understands \& searches unstructured data, Fuzzy text matching through OCR, Ontology Mining, Machine-learned ranking(MLR).\\[.5\normalbaselineskip]
     CVScan  &
    Free service, scan CV and job description and compare keywords \&amp;frequencies \&amp; match rate, includes top skills per industry (weighted).\\[.5\normalbaselineskip]
    Untapt  &
    Talent-matching based on Natural Language (not keywords), Identify future leaders based on custom data analysis, white label solution or branded, AI-driven hiring decisions.\\[.5\normalbaselineskip]
    Google Talent Solution  &
    Talent Solution uses ML technology to better understand job content and jobseeker intent, Talent Solution can interpret the vagueness of any job description, job search query, or profile search query., includes military occupational specialty code translation (MOS, AFSC, NEC).\\[.5\normalbaselineskip]
    Zoho Recruit  &
    A candidate’s match score is calculated using their skills and qualifications, contact the matched person through the platform, semantic search, radius search (location), integrates with Linkedin, parse CV, large CV database\\[.5\normalbaselineskip]
     DaXtra &
    Offered as a component deployment or hosted service, Rich structured data output, Skills taxonomy extraction, Geographical and multilingual coverage, Social media awareness, Highly accurate, Continually updated\\[.5\normalbaselineskip]
    \hline
    
  \end{tabular}

\end{table}

\subsection{Supplementary method used for matching}
There exist many platforms in the market who are already providing their service having a different business strategy. \cite{res_recomm_online} .

\begin{figure}[h!]
    \centering
  \includegraphics[width=120mm]{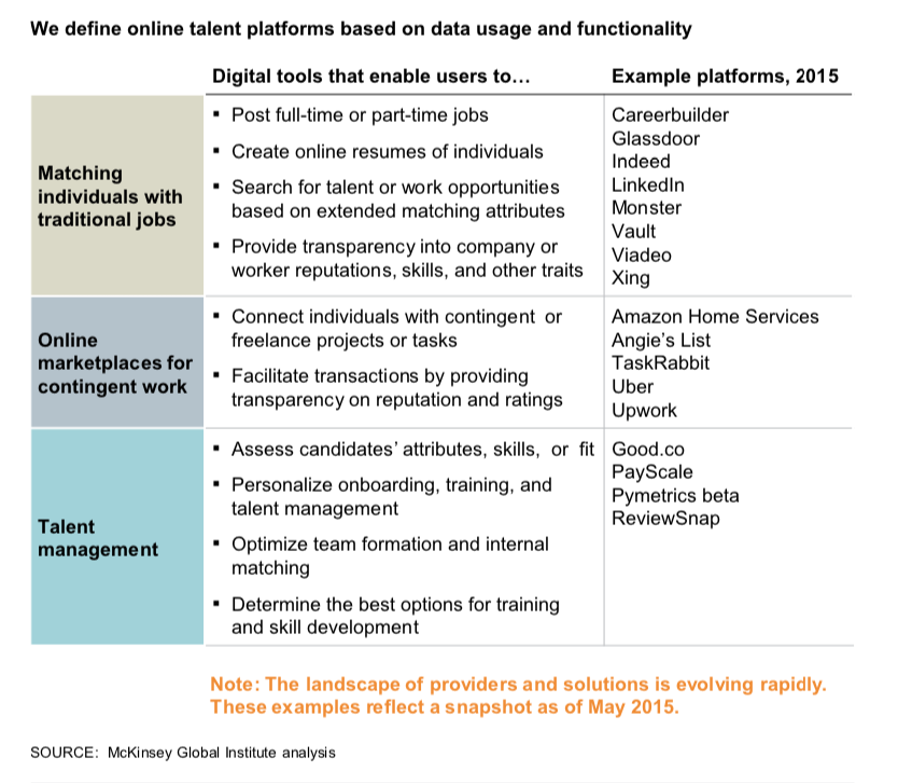}
  \caption{Existing online talent platforms}
  \label{fig:onlineplatforms_here}
\end{figure}

Existing online market tool which are providing the service to businesses in various ways as shown in Fig. \ref{fig:onlineplatforms_here}.

\subsection{General and specific aims}

\begin{figure}[h!]

\centering
\includegraphics[width=90mm]{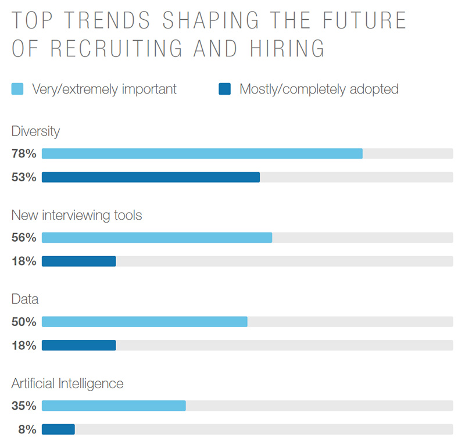}

\includegraphics[width=90mm]{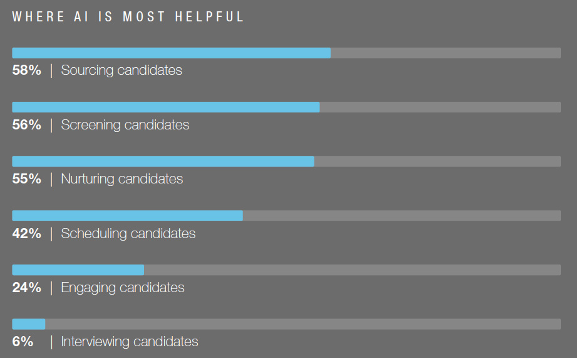}

\caption{\textbf{(a)} Four trends were identified based on numerous expert interviews and a survey of 9,000 talent leaders and hiring managers across the globe \cite{recruitment_def} \textbf{(b)} Details on hiring trends.}
\label{fig:toptrends}
\end{figure}

With the intervention of AI, the recruitment process may be completely disrupted to a new future revolution.
Fig. \ref{fig:toptrends} show the statistics of the main trends observe in the last part of section 1. Its very obvious here that the Artificial Intelligence field is still not much adopted. This is seen as an opportunity to leverage an AI in the field of HR tech industry.

Our proposed solution is to create a personalized recruitment product for an organization based on its workforce strength with which we can provide the scoring of the candidate along with the feedback (acceptance/rejection) purely driven by AI. The product which could have the ability to think multidimensional, having the ability to  take care of all the aspects between candidate and job post which can help to complete the entire recruitment process with more efficiency, effectiveness, and ultimately fit between potential candidates and recruiters.

\section{Background Research}
\subsection{What is important in a  CV ?}

As a first glance, the recruiter will be already evaluating the quality of a CV and its organization. Fact that could help or harm the overall result. Although, we won’t take into account this aspect of the CV and recruiter first encounter. Instead we’re going to extract the content and study it. 

So, for the content, CV is a structured document that can be separated into several sections. A CV could have or not each of these parts depending on the experience, the type of applicant (researcher, private/public) and simply whether or not they follow a standard structure. But whatever the CV received is, objectively, the recruiter will search for some specific parts in order to understand who the applicant is, and if he/she passes this first filter, a reading, understanding and making sense of the CV.

In order to understand the profile, an understanding of each section should be done. The different sections are: contact information, personal details, skills, professional experience, academic experience, projects, recognition and awards, publications, certifications and references. And an overall review on spelling and grammar is important too.

\subsection{Regular structure on a CV - Metadata}

As mentioned before, the list of expected input is limited, each one has its own objective on helping define the candidate’s profile.

\begin{list}{--}{}
\item Contact information: To contact the candidate.

\item Personal details like birthday, nationality, social networks, blogs or github: In order to go further the CV if desired. (further work can be web crawling to discover some traits of the applicant).

\item Skills: Have an overview of the candidate’s values and personal characteristics.

\item Professional experience: Understand the relevance of the professional path regarding the offer and the enterprise.

\item Academic experience: Extract the basis of the human capital and see if it’s pertinent and use it as an indicator.

\item Projects: Depending of the type, professional or academical, they tell about the experience or motivation 

\item Recognitions and awards: In order to differentiate from the others.

\item Publications: In a research context, their role is to describe the potential of a candidate.

\item Certifications: If necessary for the job, otherwise, a recognition

\item References: Get further feedback from the candidate, and speak about candidate’s values.

\end{list}

\subsection{Recruiter points of views}

We consulted a recruiter of Thales human resources department, who has been involved in the recruitment field in the engineering industry for more than 15 years, in order to have a better understanding of the recruitment process.

\begin{figure}[h!]
    \centering
  \includegraphics[width=120mm]{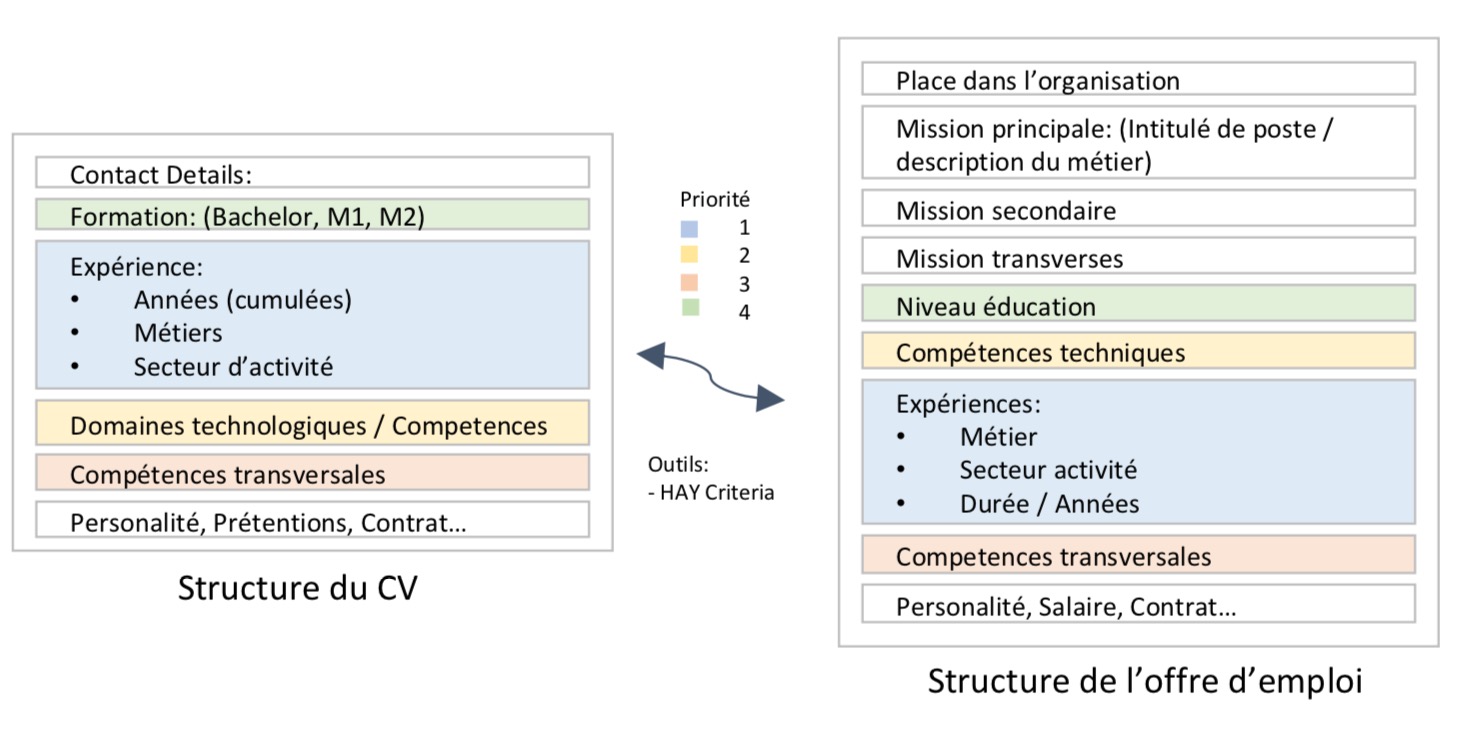}
  \caption{Basic match made by recruiter}
  \label{fig:onlineplatforms}
\end{figure}
As an initial observation, she told us that this approach was her strategy and which is used most of the time in the recruiting field.
They divide the work in two parts, extracting features from the required job posts and extracting features from the CV and then perform one to one matching between both. There are certain aspects which keep into account always for eg: School, degree.
As interesting points we find that recruiters point of view can vary, depending on the country and depending on the company. Some workers that are selected for a company, may not fit for another similar post in another company. Then, we observe a sign that the culture of a company is important to set parameters into the recruiting process.

\subsection{Related work on Parsing and Matching}

The domain of job matching has been researched since decades. AI has become talk of an hour by many researchers and  business enterprises. The researchers are creating new algorithms in the field of talent acquisition which can help the business to find the best candidates without introducing any algorithm bias\cite{resumatcher}.

In the literature, the research is sparse and not a lot of specific domain studies have been done. For instance, finding “how to parse a CV and match/recommend/class it to a job posting” is not at all available. This type of study might have been interesting because a CV is a structured document where information of different categories could be extracted and analysed in parallel to extract information and get more accurate results for each sub-structure. A similar approach to our desired project is a recommendation model by implementing a genetic algorithm that uses recruitment records to establish the users demand model \cite{resumatcher}. From the researcher perspective the matching problem has been tackled in different ways for eg:- Recommender systems are broadly accepted in various areas to suggest products, services, and information items to latent customers. Yi et al. used structured relevance models (SRM) to match résumés and jobs\cite{resumatcher}.Drigas et al. presented an expert system to match jobs and job seekers, and to recommend unemployed to the positions. The expert system used Neuro-Fuzzy rules to evaluate the matching between user profiles and job openings. they also  proposed a fuzzy logic based expert system (FES) tool for online personnel recruitment. The system uses a fuzzy distance metric to rank candidates’ profiles in the order of their eligibility for the job\cite{resumatcher}.

\subsection{Most commonly used Methodology for job matching}

The primary steps involved for a recruiters before matching a CV to a job post is understanding the job post. It is very essential for the recruiter to understand what they expect from the particular job post. In the process of which each job post is evaluated based on certain defined criteria and the candidates are accessed if they meet those criteria. The most popular defined criteria used by most of the recruiters 

\subsubsection{HAY criteria}

The HAY system is based on measuring the job against three elements which are deemed to be common in all jobs \cite{hay_guide_chart, hay_methodology}.
These elements are:
\begin{list}{--}{}
\item KNOW HOW - This measures the range of technical, planning, organising, controlling and communicating/influencing skills required in order to be able to perform the job competently.
\item PROBLEM SOLVING - This measures the degree of complexity involved in carrying out the job.
\item ACCOUNTABILITY - This measures the influence that the job has and the decisions made in achieving the end result.
\end{list}
Each job is measured against these three elements.  A numeric score for each is calculated, using charts provided by HAY Management Consultants. The total of the three scores (job units) identifies the grade into which the job falls.

\subsection{Open Source Knowledge Bases}

Many countries have made their data open source for the purpose of study on job openings, hires, and separations, providing an assessment of the availability of unfilled jobs, and information to help assess the presence or extent of labor shortages.

\subsubsection{Rome}

In France, ROME (Répertoire Opérationnel des Métiers et des Emplois) is a tool for professional mobility and the matching of offers and candidates. The ROME was built by the Pôle emploi teams with the contribution of a large network of partners (companies, branches and professional unions, AFPA...), based on a practical approach: inventory of the most common job titles/jobs, analysis of activities and skills, job grouping according to a principle of equivalence or proximity.

\subsubsection{O-NET}

The O*NET Program is the nation's primary source of occupational information. Valid data are essential to understanding the rapidly changing nature of work and how it impacts the workforce and U.S. economy. From this information, applications are developed to facilitate the development and maintenance of a skilled workforce \cite{onet_center}.

\subsubsection{ESCO}

European Skills, Competences, Qualifications and Occupations (ESCO) is a multilingual classification of European Skills, Competences, Qualifications and Occupations. ESCO is part of the Europe 2020 strategy. The ESCO classification identifies and categorises skills, competences, qualifications and occupations relevant for the EU labour market and education and training. It systematically shows the relationships between the different concepts. ESCO has been developed in an open IT format, is available for use free of charge by everyone and can be accessed via the ESCO portal.

% The table \ref{table:1} is an example of referenced elements.
 
% \begin{table}[h!]

% \centering
% \begin{tabular}{c c c}  
%  \hline
%  Section & Detail & Comments\\ [0.5ex] 
%  \hline\hline
%  Personal Information & Name (Format Names can vary) & \\ 
%   & Picture &\\
%   & Address (Country, State) &\\
%   & Picture &\\
%   & Picture &\\ [1ex] 
%  \hline
% \end{tabular}
% \caption{Table to test captions and labels}
% \label{table:1}
% \end{table}

\subsection{Strategy/Plan}
\subsubsection{Summary generated from CV’s}
Each recruiter has to search in each part and highlight the important points in the CV in order to get an overview of facts that would help him understand if the candidate is adequate for the role, and then, some that would show characteristics that most save time to the recruiter.

\subsubsection{Feedback from CV’s to the candidate}
We propose as a further step, to give feedback to the applicant about it’s CV, like if the role he’s searching for could be not very suitable for him, propose him some roles. Also, tell him where he is according to the job needs, if he should increase his skill on something else and how adequate he is in respect to similar job offers.

\subsubsection{Recommendation}
In order to do the scoring and matching we need to understand how we’re going to do it. From some research and recruiters feedback, we have come with some metrics to extract from the CV. One thing to take into account is that for several metrics, it’s existence is not certain so this fact must be taken into account. A “must have” note will be then proposed in order to mitigate the possible missing values that are not mandatory, but still, give them some importance to the fact that they exist if ever present.

\subsubsection{Stages, algorithm flow}
In order to create the final Proof of concept, we want to follow a recruiter based evaluation logic in order to optimize the processes. This would permit the flow to ignore and class CV's. So, these two are the main stages to do the recommendation.

In the case of doing the whole process of recommendation we would already know what the recruiter is searching for, so we would be able to apply the HAY job evaluation criteria in order to offer two things: drop the CV and score on relevant CV. Since the HAY criteria offers us a way to see the immediate relationship between two roles, and to understand how much the candidate's experience is adequate for the job role the company is proposing.

As a first step, "Drop". We could search for the minimum requirements the recruiter is searching for, the “must have” ones in order to do a direct match with the job posting and drop candidates who don’t have these minimal skills. Then, we would  apply the HAY criteria in order to know how much related the job position is to the experience and roles the candidate has had. If we are not able to extract this information from the candidate we would return as feedback for him to add it to the CV and reapply. If the information is “blurry”, we would simply not delete the candidate but assign a high score to the dropping criteria. Also, as a further step, we would require the recruiter feedback in order to improve this analysis when “blurry”. 

As a second step, "Score". Once all relevant profiles have been selected, we could use the HAY evaluation done as a first input to the classifying algorithm. Then, we would use the metrics in order to do a classification among the candidates. For this, we would also apply the recruiters point point of view in order to give higher or smaller scores to the metrics results.

\subsubsection{Metrics}
Table \ref{tab:table12},\ref{tab:table13},\ref{tab:table14} are the different metrics which recruiters look for while matching the CV to a job post.
% \paragraph{\color{color1}1. Professional Experience}

\begin{table}[H]
  \begin{center}
    \caption{Personal Experience}
    \label{tab:table12}
    \begin{tabular}{p{4cm} p{6cm} p{4cm}} 
      \textbf{Metric } & \textbf{Description} & \textbf{Type}\\
      \hline
      \multicolumn{3}{c|}{\textbf{Periods}}\\
      \hline
    total\_years\_of\_experience &
    from the first job to last job &
    number \\[.5\normalbaselineskip]
    experience\_occupation \_percentage &
    percentage of the time of experience that actually has been invested in working &
    number [0, 1] \\[.5\normalbaselineskip]
    experience\_shifts\_behavior &
   note describing the pause behavior between works: is it random? each time is one year? it has reduced over time? &
    number [ -1, 1 ] (don’t take into account if 0 or less) \\[.5\normalbaselineskip]
    experience\_total\_occupation \_time\_jobs\_ratio &
    ratio of time per job &
    number [0, 1]\\[.5\normalbaselineskip]
    experience\_gap\_limit\_ repetitions &
    count how many times the pauses between jobs were bigger than 9 months (tolerance + 5 days) &
    number \\[.5\normalbaselineskip]
    
      \hline
      \multicolumn{3}{c|}{\textbf{Company}}\\
      \hline
     activity\_sector &
    activity sector (civil engineering, computer science ...) &
    name \\[.5\normalbaselineskip]
    country &
    idem. &
    name\\[.5\normalbaselineskip]
    
    \hline
      \multicolumn{3}{c|}{\textbf{Activities}}\\
      \hline
experience\_action\_words \_list &
    Keep list for job matching relevance evaluation (Created,received,deployed …) &
    action\_list \\[.5\normalbaselineskip]
    experience\_important\_words \_list &
    Keep important words list for further usage in job matching (Managed Optimised Reduced Developed Increased Supported Negotiated Presented Resolved Improved...) &
    important\_list\\[.5\normalbaselineskip]
    experience\_activities\_skills\_list &
    Deduce type of activities from overall skills: management, abstraction, scientific framework.... &
    skill\_list (all types)\\[.5\normalbaselineskip]
    \hline
      \multicolumn{3}{c|}{\textbf{Role}}\\
      \hline
    experience\_role\_type &
    title of the role (would work to see the distance to the actual job position) &
    name \\[.5\normalbaselineskip]
    career\_continuity &
    where the successive roles related? &
    mapping of career sectors\\[.5\normalbaselineskip]
    \hline
     \end{tabular}
  \end{center}
\end{table}   

\begin{table}[H]
  \begin{center}
    \caption{Academics details}
    \label{tab:table13}
    \begin{tabular}{p{4cm} p{7cm} p{4cm}}
      \textbf{Metric } & \textbf{Description} & \textbf{Type}\\
      \hline
      \multicolumn{3}{c|}{\textbf{Academic Experience}}\\
      \hline
    academic\_institution\_title &
   Name &
    name \\[.5\normalbaselineskip]
    academic\_institution\_country &
    country &
    name\\[.5\normalbaselineskip]
    academic\_experience\_period &
    period &
    date\\[.5\normalbaselineskip]
    academic\_experience\_total &
    cumulated years &
    number\\[.5\normalbaselineskip]
    academic\_degree &
    degree &
    number\\[.5\normalbaselineskip]
    academic\_major &
    major &
    name\\[.5\normalbaselineskip]
    academic\_grades &
    grades &
    number\\[.5\normalbaselineskip]
    academic\_institution\_score &
    score &
    number\\[.5\normalbaselineskip]
      \hline
      \multicolumn{3}{c|}{\textbf{Academic Projects}}\\
      \hline
    aca\_project\_types &
   in acadamical purpose?, professional? entrepreneurship?] &
    name \\[.5\normalbaselineskip]
    
    aca\_project\_subjects &
   Distance measuring between role activities, and type can be done. &
    name \\[.5\normalbaselineskip]

   aca\_project\_duration\_list &
   projects duration &
    number \\[.5\normalbaselineskip]
    aca\_project\_start \_date\_lists &
   projects start date &
    date \\[.5\normalbaselineskip]
    
    aca\_project\_name\_list &
   [in acadamical purpose?, professional? entrepreneurship?] &
    name \\[.5\normalbaselineskip]
    aca\_project\_count &
   number of projects &
    number \\[.5\normalbaselineskip]
    aca\_project\_count\_if\_relevant &
   relevant projects count &
    number \\[.5\normalbaselineskip]

    aca\_skills\_list &
   skills list obtained from the project if any &
    skill\_list (all types) \\[.5\normalbaselineskip]
    aca\_action\_words &
   action words list &
    action\_list \\[.5\normalbaselineskip]
    \hline
    \end{tabular}
  \end{center}
\end{table}

\begin{table}[H]
  \begin{center}
    \caption{ Candidate Information}
    \label{tab:table14}
    \begin{tabular}{p{4cm} p{8cm} p{4cm}}
      \textbf{Metric } & \textbf{Description} & \textbf{Type}\\
      \hline
      \multicolumn{3}{c|}{\textbf{ Personal details}}\\
      \hline
    cand\_name &
  name, last name &
    name \\[.3\normalbaselineskip]
    cand\_picture &
  picture &
    blob \\[.3\normalbaselineskip]
    cand\_linkedin &
  linkedin &
    name \\[.3\normalbaselineskip]
    cand\_github &
  github &
    name \\[.3\normalbaselineskip]
    cand\_facebook &
  facebook &
    name \\[.3\normalbaselineskip]
    cand\_twitter &
  twitter &
    name \\[.3\normalbaselineskip]
    cand\_blog &
  blog page / web page &
    name \\[.3\normalbaselineskip]
    cand\_nationality &
  nationality &
    name \\[.3\normalbaselineskip]
      \hline
      \multicolumn{3}{c|}{\textbf{Contact Information}}\\
      \hline
    cand\_mail &
  mail &
    name \\[.3\normalbaselineskip]
    cand\_phone &
  phone &
    name \\[.3\normalbaselineskip]
      \hline
      \multicolumn{3}{c|}{\textbf{Skills}}\\
      \hline
    soft\_skills &
  soft skills &
    list \\[.3\normalbaselineskip]
    transversal\_skills &
  transversal &
    list \\[.3\normalbaselineskip]
    language\_skills &
  languages &
    list \\[.3\normalbaselineskip]
      \hline
      \multicolumn{3}{c|}{\textbf{Recognitions/awards}}\\
      \hline
    award\_type &
  type &
    name \\[.3\normalbaselineskip]
    award\_year &
  year &
    number \\[.3\normalbaselineskip]
    award\_name &
  name &
    name \\[.3\normalbaselineskip]
%       \hline
%       \multicolumn{3}{c|}{\textbf{Score (normalized)}}\\
%       \hline
%     award\_total &
%   count all of them &
%     number \\[.5\normalbaselineskip]
%     award\_freq &
%   form start to beginning, divided by the total &
%     number \\[.5\normalbaselineskip]
%     award\_behavior &
%   has it been constant? is it increasing? is it decreasing? &
%     [-1,1] \\[.5\normalbaselineskip]
      \hline
      \multicolumn{3}{c|}{\textbf{Publications}}\\
      \hline
    pub\_type &
  type/subject &
    name \\[.3\normalbaselineskip]
    pub\_year &
  year &
  number\\[.3\normalbaselineskip]
    pub\_name &
  name &
    name \\[.3\normalbaselineskip]
    pub\_magazine &
  magazine &
    name \\[.3\normalbaselineskip]
    pub\_impact &
  impact [local, national, international] &
    name \\[.3\normalbaselineskip]
    pub\_coworkers &
  coworkers &
    name \\[.3\normalbaselineskip]
      \hline
      \multicolumn{3}{c|}{\textbf{Certifications}}\\
      \hline
    cert\_type &
  type (match if any certification needed &
    name \\[.3\normalbaselineskip]
    cert\_name &
  name &
  name\\[.3\normalbaselineskip]
  cert\_year &
  year &
    number \\[.3\normalbaselineskip]
    cert\_date\_validity &
  until year &
    number \\[.3\normalbaselineskip]
      \hline
      \multicolumn{3}{c|}{\textbf{References}}\\
      \hline
    ref\_job &
  correspondent job &
    name/number \\[.3\normalbaselineskip]
    ref\_tone &
  tone (negative, positive, can’t say) &
  {-1,0,1}\\[.3\normalbaselineskip]
  ref\_match\_job &
  correspondence with activities &
    {1,0} \\[.3\normalbaselineskip]
    ref\_skills &
  skills &
    list \\[.3\normalbaselineskip]
    ref\_name &
  name &
    name \\[.3\normalbaselineskip]
    ref\_contact\_info &
  contact info &
    name \\[.3\normalbaselineskip]
      \hline
      \multicolumn{3}{c|}{\textbf{Candidate’s summary}}\\
      \hline
    content &
  summary, very variate, not structured, for the moment, just identify it, and pass it as it is to the recruiter &
  name \\[.3\normalbaselineskip]
  \hline
    \end{tabular}
  \end{center}
\end{table}
% table to be added here 

\subsection{Ontology}
Leveraging the domain of NLP to build a HR ontology which consists of thirteen modular ontologies : competence, education, job offer, job seeker,  language, occupation, skill and Time can play a very important role.  The main sub ontologies are the job Offer and job Seeker, which are intended to represent the structure of a job posting and CV respectively. While these two sub ontologies were built taking as a starting point some HR-XML \cite{hr_xml} recommendations, the other sub ontologies were derived from the available international standards (like NACE, ISCO-88 (COM), FOET, etc.) and ES classifications and international codes (like ISO 3166, ISO 6392, etc.) that best fit the European requirements.

\begin{figure}[H]
    \centering
  \includegraphics[width=120mm]{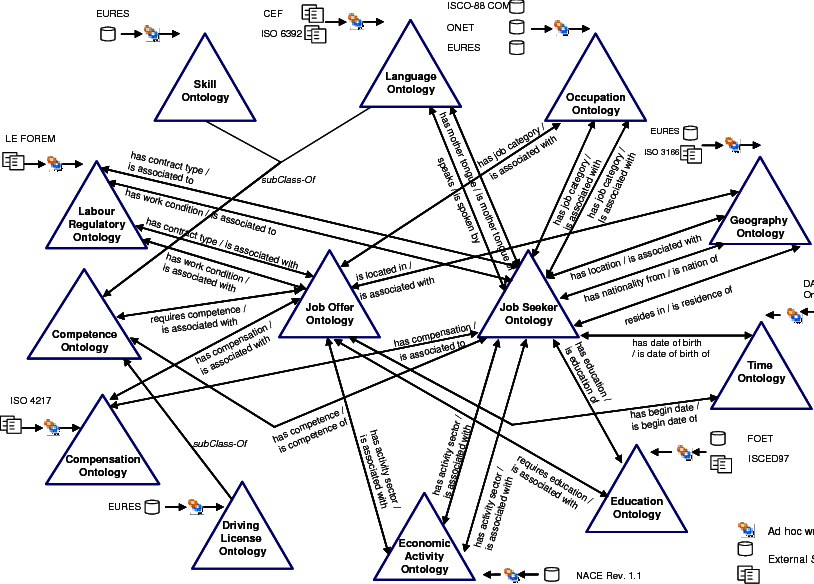}
  \caption{Main ad-hoc relationships between the modular ontologies. }
\end{figure}

Details of the ontology is explained well in “Reusing Human Resources Management Standards for Employment Services” \cite{hr_mgmt_stdrs}.

In the scope of our project we build a basic job skill ontology based on online available technology, to build an ontology we intend to build these ontologies using the reference provided in \cite{taxo_skills_job_ad}. The flow chart shown below depicts how to build basic taxonomies which can be further converted into ontologies.

\begin{figure}[h!]
    \centering
  \includegraphics[width=110mm]{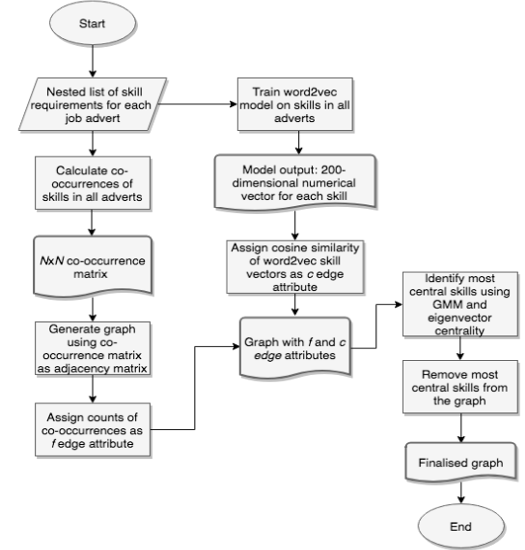}
  \caption{Flow chart for building and preparing the skills graph}
\end{figure}

\begin{figure}[H]
    \centering
  \includegraphics[width=110mm]{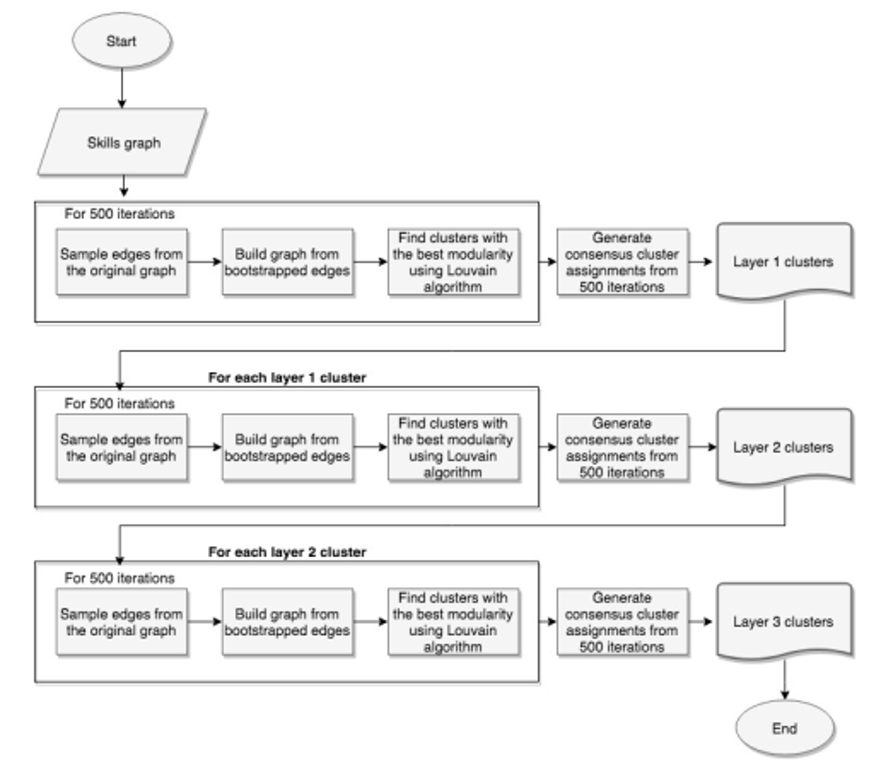}
  \caption{ Flow chart for detecting hierarchical communities in the skills graph}
\end{figure}

\subsection{ Model Proposition (Input/Output)}
\subsubsection{No data model proposal}
Approach to take: divide and conquer plus expertise dependence. Since the recruitment process is singular for each sector and even for each enterprise or recruiting framework, try to do a general approach remains exhaustive and even impossible to tune. An alternative to attack this issue is to let the recruiting area of the enterprise (our final customer) the freedom to tune some metrics that would be introduced to our final algorithm. This would permit a broader impact in the market. since, behind the scenes, the algorithm would remain the same without need for us to adapt it trying to handle the biggest amount of study cases. Of course, this could lead to a “difficult product” so the tuning parameters should remain reduced. This would remain the more delicate “issue” for the customer's point of view since with a data based model, they would not need to do any tuning, so the objective would be to enlighten its possibilities and advantages.

In order to solve each small challenge, each of the metrics will be taken into account. A “very important”, “important” and “not important” category would be proposed for the main categories and maybe for each category if the recruiter needed it. Thus, he could adapt in order to comply with the needs of the role to be fulfilled, including all the cultural and organizational characteristics that the company must be seeking.

\begin{figure}[H]
  \includegraphics[width=150mm]{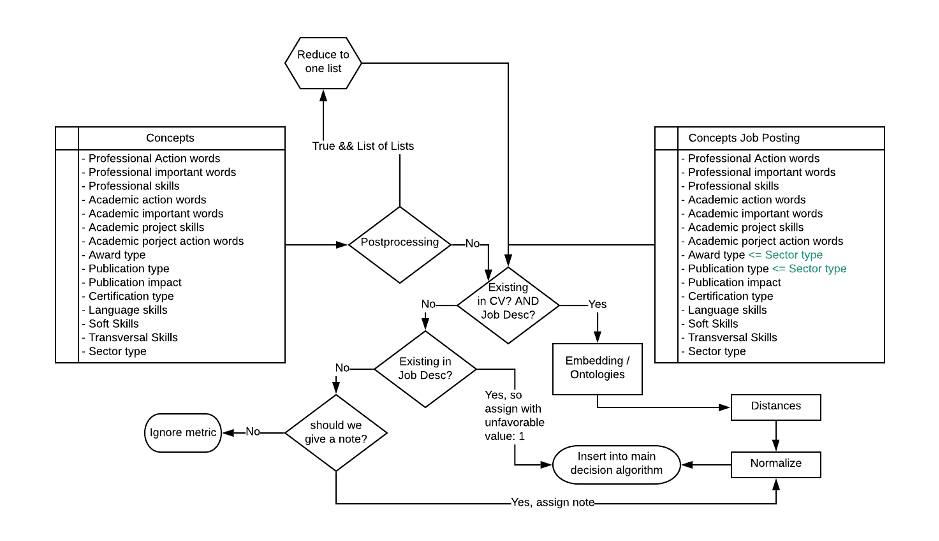}
  \caption{ Example, flowchart for concepts treatment}
\end{figure}

As explained in the figure above, each type of metrics { numbers, names, concepts } would be treated with an algorithm to extract value that could be inserted into the “main algorithm” that would help to score. 

First, a filter would be done so there would be a list that when found, extracted and matched less than required, the treatment of CV would stop by justifying the reason so that it still appears in the list of candidates’ CV's but as “not qualified, reason: X”. 

Next, the extraction of these metrics would continue and contribute to feed the “main algorithm”. The “main algorithm”model builds a skill graph by using ontology in order to do the match between CV and job post. The key concepts are identified from the text using text processing and are matched with the nodes of the ontology in order to get detailed information related to the identified concept. In addition we can take any additional requirements from the organization which can enhance their work culture. For eg if the organization emphasizes much on soft skill, creativity and  other culture fit criteria needed within. The algorithm will exploit to match every section extensively like experience ,skills, soft skills, academic projects, motivation. For eg for the skill section ,it will generate a skill graph for both the job post and CV and measure the similarity between them.  And finally, a multi criteria approach.  such as MR sort to obtain a final score.

\section{Implementation}

\emph{For the implementation, we propose the minimal viable product, So as discussed in  \textbf{Model Propositions} , we managed to get existent ontologies and embedding, to be able to work out functionalities related to the proposed model. Since building an ontology exclusively for every section is a time taking process. Here, we focus on matching technical skill along with cultural feature between CV and job post.
Functionalities like the parser, and more dimension evaluations had to be ignored in order to complete a first functional product.  Moreover, data was created by using an existing software that organized already all CV structured data so that we could exploit it directly.}

% So, next chapters correspond to the development of what we call the "first iteration".

\subsection{Create ontology's}
The structure of ontologies borrows a lot from graph theory, and For instance, when considering competencies, each competency is a ‘node’ and each relationship between competencies is an
‘edge’. Ontologies are represented as undirected graphs .

In order to create an ontology for the skill development, Based on the research which we made on the online available ontologies. We chose to work with CSO Ontology. We also created manually the domain specific ontology from the crawled job posts.

\subsubsection{Technical Skill ontology}

The Computer Science Ontology (CSO) is a large-scale ontology of research areas that was automatically generated using the Klink-2 algorithm on the Explore dataset, which consists of about 16 million publications, mainly in the field of Computer Science [https://cso.kmi.open.ac.uk/home ].The Klink-2 algorithm combines semantic technologies, machine learning, and knowledge from external sources to automatically generate a fully populated ontology of research areas.It also includes Linguistics, Geometry. The current version of CSO includes 26K topics and 226K semantic relationships.

 It includes five semantic relations:
 \begin{list}{--}{}
 \item relatedEquivalent, which indicates that two topics can be treated as equivalent for the purpose of exploring research data (e.g., Ontology Matching and Ontology Mapping). For the sake of avoiding technical jargon, in the CSO Portal this predicate is referred to as alternative label of
 
 \item skos:broaderGeneric, which indicates that a topic is a super-area of another one (e.g., Semantic Web is a super-area of Linked Data). This predicate is referred to as parent of in the portal. The inverse relation (child of) is instead implicit
 
 \item contributesTo, which indicates that the research output of one topic contributes to another. For instance, research in Ontology Engineering contributes to Semantic Web, but arguably Ontology Engineering is not a sub-area of Semantic Web – that is, there is plenty of research in Ontology Engineering outside the Semantic Web area. 
 
 \item rdf:type, this relation is used to state that a resource is an instance of a class.For example, a resource in our ontology is an instance of topic.
 
 \item rdfs:label, this relation is used to provide a human-readable version of a resource’s name.

 \end{list}
 
 \begin{figure}[h!]
    \centering
  \includegraphics[width=150mm]{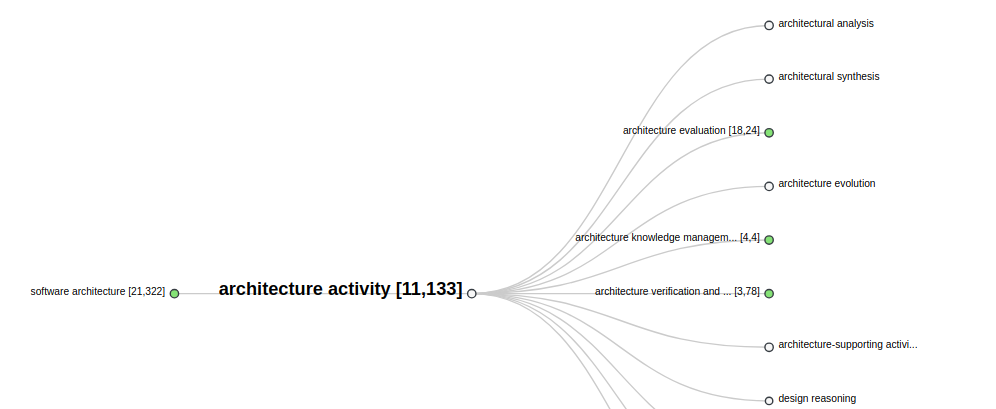}
  \caption{CSO ontology overview }
\end{figure}

\subsubsection{CSO generation}

The working of Klink2 algorithm \cite{klink} takes as input a set of keywords and investigates their relationship with the set of their most co-occurring keywords. The algorithm tries to find the semantic relationship between keyword x and y by the means of three metrics which are  hierarchical relationship, temporal relationship and similarity . The first two used to to detect skos:broaderGeneric and contributesTo relationships, while the latter is used to infer relatedEquivalent relationships.

\subsubsection{Domain Skill ontology}

In order to create a domain skill ontology, we collected the job posts of the particular domain for e.g we focused on creating an ontology in the domain of data science. As it helps to find the key terms which exist related to the domain in such a job post. For example the cso ontology lag the term such as algorithms or tools which are explicitly related to particular domain. Here we build a hierarchical based ontology where the nodes of the same type have some special semantics for defining parent/child relationships as this is a very common relationship necessary to express existing child and parent relationship frameworks. A node defined as a parent generally is a broader version of all of its children, having many shared attributes. For instance, ESCO defines an ‘advanced nurse practitioner’ and ‘specialist nurse’ as both being children of ‘nursing professionals’. These occupations understandably share many competencies, and it is easy to imagine experience in any type of nursing professional occupation as being broadly applicable to other nursing professional occupations. This parent/child hierarchy is necessary, but not itself sufficient for defining rich ontology is capable of expressing the relationships between other nodes or child.

\subsubsection{Domain skill ontology generation}

We created a large text corpus where we collected all the job posts in the data science skill domain. The job post was crawled from the Dice platform (https://www.dice.com/) . We collected 10000 job posts. To generate the ontologies we employ machine learning methods, such as word embedding and clustering algorithms.

All the stop words existing in the job post corpus are removed and then we created a concept based on the number of occurrence of words in the whole corpus based on n-gram technique. We used the library word cloud \footnote{$https:\//github.com\//amueller\//word_cloud$} which chose the top 200 originated concept. After creating the vectors using the nltk vectorize model, the cluster is being formed using the K-Mean Algorithm. After exploring the obtained cluster thoroughly we create a basic ontology using the protege software \footnote{https://protege.stanford.edu}. The ontology can be accessed using \url{http://owlgred.lumii.lv/online\_visualization/lli4/}.
\newline
Figure 8 shows the brief overview of domain data science ontology.

\begin{figure}[h!]
    \centering
  \includegraphics[width=150mm]{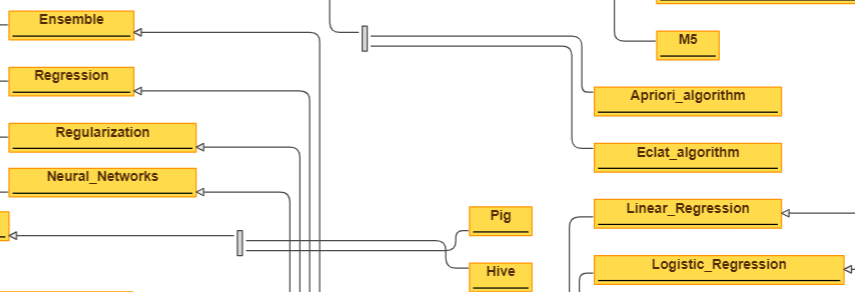}
  \caption{Data science domain skill ontology }
\end{figure}

\subsubsection{Cultural values ontology}

In order to understand the culture that a text could explain by itself from an enterprise point of view we approached theory that studies this as concepts that permit a classification of the enterprises on a cultural level. So to start with, to locate this problem, we're going to make some remarks: we're treating CVs and job postings, so general and domain specific terms would be found, and the context happens in an enterprise-like environment, an organization, whether public or private. These kinds of documents are not very explainable texts since they don't have a lot of complete phrases that would need a deeper sense analysis to get better results. This being said, we're going to explain how we managed to approach the extraction of what we call "organizational culture" from CVs and job postings.

To start with, we  based our understanding in proposed organizational culture understanding theory \cite{org_cult} where several researchers propose a way of characterizing an organization. We use a global view of the characterization. As a first step, four levels can characterize culture in an organization \{\textit{Symbols, Heroes, Rituals, Values} \} from which we're just interested in the values since they are only ones that can be extracted from text. The other levels imply abstraction or more inside-company behaviors and traditions that can't be forcefully extracted from CV or a job posting. As a second step, values have been distinguished in six different concepts \{ \textit{Power Distance, Individualism, Uncertainty Avoidance, Masculinity \& Femininity, Long Term Orientation, Indulgence Vs Restraint} \} and each of these concept is subdivided in two "antonym" set of values describing its parent \textit{(showed in the next table)}.

\begin{table}[h!]
  \caption{Organizational Culture Dimensions}
  \centering
  \begin{tabular}{p{4cm} p{4cm}} 
    \hline\hline 
    Organizational Culture Dimension & Concepts Comparison \\
    %heading
\hline
    Power Distance &
    Small $\leftarrow$ - $\rightarrow$ Large \\
    Individualism &
    Individualism $\leftarrow$ - $\rightarrow$ Collectivism \\
    Uncertainty Avoidance &
    Weak $\leftarrow$ - $\rightarrow$ Strong \\
    Masculinity \& Femininity &
    Masculinity $\leftarrow$ - $\rightarrow$ Femininity \\
    Long Term Orientation & 
    Short $\leftarrow$ - $\rightarrow$ Long \\
    Indulgence Vs Restraint &
    Indulgence $\leftarrow$ - $\rightarrow$ Restraint \\
\hline %inserts single line
\end{tabular}
\label{table:nonlin} % is used to refer this table in the text

\end{table}

Each of these "antonym" concepts has other concepts that describe it (for more information see \cite{org_cult}). For example, for Power Distance, we have the next descriptors: \textit{decentralization/centralization, management by experience/management by rules, autonomy of employee/order directed employee, pragmatic superior relationships/emotional superior relationships, no privileges/privileges}. 

Thus, in order to make this information useful from a practical point of view, and create CV and job posting profiles at the cultural level, we proposed a graph. The approach was to develop a directed graph that would handle a multiple level division of concepts until the very end where terms would describe concepts following the next logic: {Culture -$>$ Values -$>$ Organizational Culture Dimensions -$>$ Concepts Comparison ("antonyms") -$>$ Concepts Definition Concepts (descriptors) -$>$ terms (descriptors' terms)}. So, a tree-like graph where each of these descriptors has terms that describe them, let's call them "descriptors' terms". These descriptors' terms were proposed by us, so improvements can be made. The way to assign the terms was to do a limited and definition directed list of terms referring to the parent node searching for definitions and extracting the most coherent and related terms.

\begin{figure}[H]
    \centering
  \includegraphics[width=150mm]{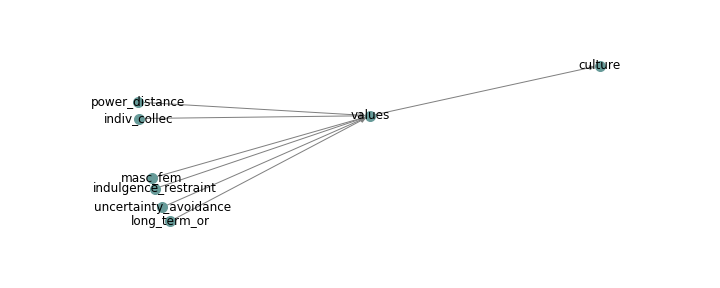}
  \caption{3 First levels of culture graph}
\end{figure}

Furthermore, the idea to use a directed graph with no inter-related terms or concept resides in a practical decision where simplicity and the nature of the task intervenes. Since the task, "extract culture profile", means that the vocabulary to use is general and not domain specific, already trained models could be implemented. To this end, an embedded model has been chosen: the GloVe model \cite{glove}. This model was chosen against the word2vec model because of four reasons. The first one is because it was stated that somehow it maintains a better analogy. The second one, by doing some tests, the results given by simulating the application of terms (as it would happen with CV's and job postings) threw congruent and meaningful terms. And the third one, word2vec model was trained in a set of news texts (google News), so could be news context specific, and glove in Wikipedia corpus, so a broader set of contexts. And fourth, because when searching for similarities, of terms against a set of words we want to be antonyms by less similar and for the word2vec model, the antonyms used to be more similar than in the GloVe model. Still, this model could be changed, simply, it has to respect the \textit{gensim \cite{gensim} word2vec} standards.

\begin{table}[H]
  \centering
  \begin{tabular}{p{4cm} p{4cm}} 
    \hline\hline 
    GloVe & Word2Vec \\
    %heading
    \hline
    'centralised', 'decentralized', 'hierarchical', 'decentralised', 'bureaucracy' & 'decentralized', 'centralizing', 'centralize', 'Centralized', 'centrally\_managed' \\
    \hline %inserts single line
    \end{tabular}
    \label{table:nonlin} % is used to refer this table in the text
     \caption{Example of results similar to "centralized"}
\end{table}

Finally, now that the structure of the graph and the why's of the structure have been answered, we can conclude by saying that the idea for this ontology is simply a term list (descriptors' terms) that represent concepts (descriptors). These concepts are sets of concepts (descriptors) that are part of a main concept definition and the ensemble of these main concept definitions are somehow antonyms that belong to wider concepts (organizational culture dimensions) which describe the values of the culture. At the end the part of the graph that will participate in the matching will be the leaves (descriptors' terms).

\begin{figure}[h!]
    \centering
  \includegraphics[width=130mm]{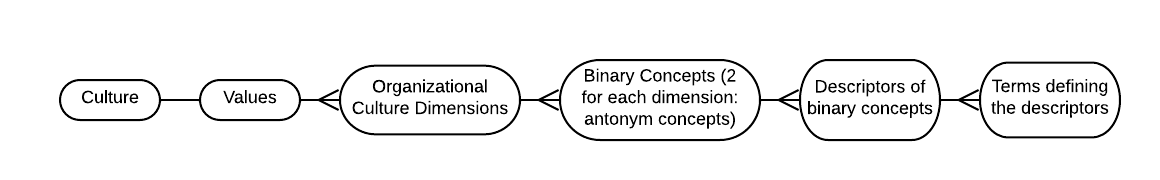}
  \caption{Structure of Cultural Graph}
\end{figure}
\subsection{Matching}

In order to do fair matching between the job post and CV'S (github code \cite{Rudresh2020}). We create a graph with the help of the above created ontologies for both job post and CV for each section(general skill, domain skill and culture).

Similarity matrix is calculated between each section graph obtained from both the job post and CV. The obtained matrix is normalized to a matching score. The library used to measure the similarity between the two graphs is known as GMatch4py. GMatch4py follows the algorithm of graph edit distance by combining Hausdorff matching and greedy assignment \cite{GMatch4py}. After we receive the score from the different sections such as General skill match, domain skill match  and cultural match. we aggregate it to the common score using MR sort algorithm.

\subsubsection{Creating Skill Graph from Ontologies}

In order to create a graph the algorithm takes the job description or the candidate work experience text as the input  and outputs a list of relevant concepts from the job and CV's. For the Skill graph generation we followed the similar approach followed by CSO classifier \cite{CSO}.

\begin{figure}[h!]
    \centering
    \includegraphics[width=150mm]{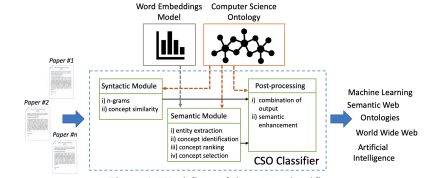}
  \caption{Workflow of CSO Classifier }
\end{figure}

It consists of two main components: (i) the syntactic module and (ii) the semantic module. 

The syntactic module parses the input documents and identifies concepts that are explicitly referred to in the document. The semantic module uses part-of-speech tagging to identify promising terms and then exploits word embeddings to infer semantically related topics. Finally, the graph combines the results of these two modules and enhances them by including relevant super-areas.

\paragraph{Syntactic Module}

The syntactic module maps n-gram chunks in the text to concepts. The algorithm removes the stop words and collects the unigrams, bigrams and trigrams chunks. Then for each n-gram, it computes the levenshtein similarity with the labels of the topic in ontologies. the minimum similarity level can be set manually and it has been set to 0.94. This value allows us to recognize many variations between concept and ontologies.

\paragraph{Semantic Module}

The semantic module was designed to find topics that are semantically related to the text. These topics are explicitly not mentioned in the text. Here it requires the word embeddings produced by word2vec to compute the semantic similarity between the terms in the text and the ontologies.
.

it follows four step .
\begin{list}{--}{}

\item  Entity extraction.
\item  Ontology concept identification.
\item  Concept ranking.
\item  Concept selection.
\item  Combined generated graph

\end{list}

The word embedding model was created by CSO using the word2vec model. The model is trained on text collected from the technical research paper in the domain of computer science.

\paragraph{Entity extraction}

The concepts can be represented either by nouns or adjectives followed by nouns. The classifier tags each word according to its part of speech (e.g., nouns, verbs, adjectives, adverbs) and then applies a grammar-based chunk parser to identify chunks of words, expressed by the grammar.

\paragraph{Concept identification}

The extracted concepts in the entity extraction stage are decomposes further into n-grams. Then similarity is measured between  the n-grams with the ontology. The scores with the top 10 similar words are identified as the concepts.

\paragraph{Concept ranking}

Since it's possible that the above step may develop a lot of topics from the ontology with the help of n grams similarity to the nodes in which there may be the concepts which might not be related to the topic we are dealing with. It means many of the identified topics could be unrelated. So in order to choose the concept which is a really important relevance score is calculated which is the product between the number of times it was identified (frequency) and the number of unique n-grams that led to it (diversity).  If a concept is directly available in the ontology, its score is set to the maximum score.

\paragraph{Concept selection}

Once the relevance score is identified for all the generated topics. The topics are plotted distributionally and the elbow  method is implemented in order to implement the top relevant topic which could be helpful in order to do the matching. \cite{elbow}

\paragraph{Combined generated graph}

The obtained topics from the both the semantic and syntactic modules are combined together. It then explored the topics by inferring all their direct super topics, exploiting the superTopicOf relationship within ontology. For instance, when the classifier extracts the topic “machine learning”, it will infer also “artificial intelligence”. All the identified returns by both the modules are  stored in the dictionary and it is further converted into graphs with the help of networkx library.

\begin{figure}[h!]
    \centering
  \includegraphics[width=150mm]{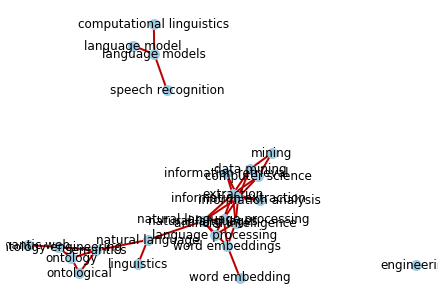}
  \caption{Generated Skill graph from job post/CV }
\end{figure}

\subsubsection{Cultural Match}

As a reminder, to describe or understand the cultural profile we have Organizational Culture Dimensions that have two antonym concepts which have several descriptors and each those have terms. So, in order to do the matching a profile of the CV or job posting is done with the help of the cultural graph, so each text will have a cultural graph profile. And then these profiles will be compared.

In order to better understand, we will explain the procedure by steps:
\begin{itemize}
    \item Calculate the cosine similarities between the descriptors' terms and the text (this is done by using the embedded model)
    \begin{itemize}
        \item \textit{ex: decentralized organization = 0.62, centralized organization = 0.52}
    \end{itemize}
    \item The mean of the similarities values of the descriptors' terms will be stored in each of the antonym concepts, this way, we'll know the belonging of each of the antonym concepts to the text, or, in other words, we will have the text profiled in the cultural graph. 
    \begin{itemize}
        \item \textit{ex: small power distance = 0.86, large power distance = 0.56}
    \end{itemize}
    \item So, we can obtain several cultural graphs that profile different CV's or job postings 
    \item To compare an euclidean distance between both profiles would be measured.
\end{itemize}

This is how the similarity between a CV (or several CV's) and job posting can be done.

\subsubsection{Education Match}

For the education match between the job and CV’s. The sovren parser parses the degree and the name of the school from the job post and CV's. Lookup dictionary is created with all the equivalent degrees related to a particular degree. For eg MSc, master, BAC+5 belongs to the same category. This match is done by keyword match in the dictionary where the required degree from the job post is searched in the degree obtained in candidate CV. 

If the candidate required degree is inferior to the degree required in the job post. The candidate gets rejected and is not processed for further stages.

\subsubsection{Required Skill Match}

All the required skills parsed from the job post are collected and is matched to the Skill/Concept found in the candidate skill graph. Based on the number of skills from the job post matched in obtained skill from candidate CV  scored is assigned. For eg if out of 4 skill 3 skill is matched in candidate CV, then the  calculated score is 0.75.

\subsection{Multiple CVs to Job Post matching}

One task is to match a CV with a job post, where the thing to do is to somehow compare different points of view of the texts such as culture, domain, education, and others. And another task to accomplish is to compare several CVs to a job posting, a task that actually would be the most common used by recruiters in order to understand the match of a candidate's profile with the job post a recruiter is promoting. In order to do this we will serve ourselves in two steps process: \textit{filtering and matching}.

\subsubsection{Filtering}

In order to accomplish this task, we will use the theory seen in HAY criteria, where there are some requirements that must be met and others that could be not mandatory. 

As one of the proposals, the user would be able to "tune" the mandatory fields in order to implement by himself this filtering process. This could be done as a second iteration over the proposed solution since for now, the proposed solution has limited analysis axes and so, the development of this part would be excessive compared to the actual functionality. For the moment, as a first approach, some filtering concepts have been imposed and CV's would be filtered taking them into account.

So, as per this iteration, simply the education axis will be taken into account. So, a comparison of required education level against the actual education level will be made and just the ones that meet or exceed the requirements will be passed.

For this end, sovren software helps a lot by telling us already the required skills that a job posting has.

\subsubsection{Sorting}

For the matching part, the help of a multi criteria algorithm is used in order to accomplish the task. When having several CV’s, even hundreds, the help of a sorting algorithm may be very cherished. So, how to sort? As recruiters may have it's own idea in mind of the aspects they want to favor depending on the situation, the sector and the needs of the company, we propose a user friendly sorting by letting the recruiters to "tune" the parameters. 

The parameters could be highly tuned in different aspects if the user needed it. Meaning that for each main axis of study, different sub parameters could be adjusted according to needs. For example, if there was a team that needed someone with a soft and friendly character because it's full of strong and difficult characters (true story), the recruiter should be able to tune this part of the sorting algorithm. But, mainly, the user should be able to tune five or maximum seven dimensions since, as per research and recommendation, those are the maximum quantity of dimensions that a person can handle. So, even if the tuning could be expanded, as a first proposal, seven is our maximum and the inputs will be directly asked to the user.

The sorting dimensions to take into account will be integers between zero and three included with the meanings: \{ 0-not interested, 1-poorly interested, 2-interested, 3-very interested \}. In this way, the user can express his interest in each of the axes in a scale from zero to three. The algorithm used is the multi-criteria majority-rule sorting algorithm \cite{mrsort_pm}.

This way the user is able to tune the dimensions indicating which interest him the most, having $n^{n-1} * (n-1)$ different combinations to adjust, being the number of dimensions to study. In our cases, we're proposing for the moment, and as part of the first iteration four dimensions { skills, domain skills, culture, required skills }.

\section{Evaluation}

In order to evaluate our system, we have implemented two different use cases. In the first use case we test the function \textbf{ManyToOne matching} that allows to filter CV's and give a score to the selected ones. The second use case evaluates the function \textbf{OneToOne matching} that allows to see the level of correspondence between one CV and one Job post.

The CV's used to evaluate this tool were downloaded from the Naukri database \cite{freesearchNaukri} using the following filters: "Search by Keywords: data", "Total Experience: 0 to 2 years'' and "Candidate Age: 20 to 30 years". The job posts used were obtained from Linkedin. The search for these job posts focused on a data science internship with 0 to 2 years of experience. Both the CV's and Job post were parsed using the demo version of the sovren parser tool \cite{sovrentool}. We collected 120 CV's and 11 job posts. This data is stored in our Mongo database.

For this evaluation, we have selected 5 different CV's which include different profiles. We can define them as technical and business oriented profiles. We have also selected two different job posts, technical and a business oriented one as well. The test dataset was the following:

\begin{table}[h!]
  \centering
  \begin{tabular}{p{1cm} p{4cm}} 
    \hline\hline
    ID & MongoID\\
    %heading
\hline
    CV1 & 5e60f5895a90883323e38bcc \\[.5\normalbaselineskip]
    CV2 & 5e60f58a5a90883323e38bdc \\[.5\normalbaselineskip]
    CV3 & 5e60f58a5a90883323e38bcf \\[.5\normalbaselineskip]
    CV4 & 5e60f58b5a90883323e38bf1 \\[.5\normalbaselineskip]
    CV5 & 5e60f58a5a90883323e38bdb \\[.5\normalbaselineskip]
    \hline 
  \end{tabular}
 \end{table}
 
 \begin{table}[h!]
  \centering
  \begin{tabular}{p{3cm} p{6cm} p{6cm}} 
    \hline\hline 
    ID & MongoID & Job Offer\\
    \hline
    BusinessOrientedJob & 5e60f5895a90883323e38bcc &
    Intern Data Science - Product Analytics at Criteo
    \\[.5\normalbaselineskip]
    TechnicalOrientedJob & 5e64cbef837ba015d90abc79 &
    Intern Data Science at Multivac
    \\[.5\normalbaselineskip]
    \hline
  \end{tabular}
 \end{table}

\subsection{Use case 1 - One To Many Matching}

\subsubsection{Result from test 1: 5 CVs and Business Oriented Job Post} 
\begin{list}{--}{}
\item  \textbf{Input weights to define the priority of sections:} DomainSkillsMatch: 2, SkillsMatch: 2, CultureMatch: 2

\begin{table}[h!]
  \centering
  \begin{tabular}{p{1cm} p{3cm} p{2cm} p{2cm} p{2cm} p{2cm}} 
    \hline \hline 
    ID & DomainSkillsMatch & SkillsMatch & CultureMatch & MRValues \\
    \hline 
    CV1 & 1.428 & 1.420 & 0.944 & 0.743 \\[.5\normalbaselineskip]
    CV2 & 1.415 & 1.418 & 0.913 & 0.739 \\[.5\normalbaselineskip]
    CV3 & 1.420 & 1.417 & 0.889 & 0.736 \\[.5\normalbaselineskip]
    CV4 & 1.427 & 1.431 & 0.845 & 0.730 \\[.5\normalbaselineskip]
    CV5 & 1.414 & 1.419 & 0.828 & 0.728 \\[.5\normalbaselineskip]
    \hline 
  \end{tabular}
 \end{table}
\newpage
 \item  \textbf{Input weights to define the priority of sections:} DomainSkillsMatch: 3, SkillsMatch: 3, CultureMatch: 1
 \begin{table}[h!]
  \centering
  \begin{tabular}{p{1cm} p{3cm} p{2cm} p{2cm} p{2cm} p{2cm}} 
    \hline \hline 
    ID & DomainSkillsMatch & SkillsMatch & CultureMatch & MRValues \\
    \hline 
    CV1 & 1.428 & 1.420 & 0.944 & 0.774 \\[.5\normalbaselineskip]
    CV2 & 1.415 & 1.418 & 0.913 & 0.772 \\[.5\normalbaselineskip]
    CV3 & 1.420 & 1.417 & 0.889 & 0.771 \\[.5\normalbaselineskip]
    CV4 &  1.427 & 1.431 & 0.845 & 0.769 \\[.5\normalbaselineskip]
    CV5 & 1.414 & 1.419 & 0.828 & 0.768 \\[.5\normalbaselineskip]
    \hline 
  \end{tabular}
 \end{table}
\end{list}

\subsubsection{Result from test 2: 5 CVs and Technical Oriented Job Post}

\begin{list}{--}{}
\item  \textbf{Input weights to define the priority of sections:} DomainSkillsMatch: 2, SkillsMatch: 2, CultureMatch: 2
\begin{table}[H]
  \centering
  \begin{tabular}{p{1cm} p{3cm} p{2cm} p{2cm} p{2cm} p{2cm}} \hline \hline 
    ID  & DomainSkillsMatch & SkillsMatch & CultureMatch & MRValues \\
    \hline 
    CV3 & 1.418 & 1.432 & 0.940 & 0.743 \\[.5\normalbaselineskip]
    CV4 & 1.427 & 1.426 & 0.928 & 0.741 \\[.5\normalbaselineskip]
    CV5 & 1.414 & 1.420 & 0.924 & 0.740 \\[.5\normalbaselineskip]
    CV2 & 1.414 & 1.417 & 0.891 & 0.736 \\[.5\normalbaselineskip]
    CV1 & 1.429 & 1.435 & 0.849 & 0.731 \\[.5\normalbaselineskip]
   \hline 
  \end{tabular}
 \end{table}
 
 \item  \textbf{Input weights to define the priority of sections:} DomainSkillsMatch: 3, SkillsMatch: 3, CultureMatch: 1
\begin{table}[H]
  \centering
  \begin{tabular}{p{1cm} p{3cm} p{2cm} p{2cm} p{2cm} p{2cm}} \hline \hline 
    ID  & DomainSkillsMatch & SkillsMatch & CultureMatch & MRValues \\
   \hline 
    CV3 & 1.418 & 1.436 & 0.940 & 0.774 \\[.5\normalbaselineskip]
    CV4 & 1.429 & 1.434 & 0.928 & 0.774 \\[.5\normalbaselineskip]
    CV5 & 1.414 & 1.420 & 0.924 & 0.773 \\[.5\normalbaselineskip]
    CV2 & 1.414 & 1.417 & 0.891 & 0.772 \\[.5\normalbaselineskip]
    CV1 & 1.432 & 1.443 & 0.849 & 0.769 \\[.5\normalbaselineskip]
   \hline 
  \end{tabular}
 \end{table}
 \end{list}
 From these results we can observe how CV3 CV4 and CV5 are more likely to be selected for a technical job offer and CV1 and CV2 are more likely to be selected for a business oriented job.
 
 We also observe that changing the weight to evaluate each section, is not changing the sorting of the CVs, however the MR score does change.
 
\subsection{Use case 2 - One To One Matching}
Since CV1 and CV3 were the CV's with great scores in both cases. Let's analyze why they have this result.

\begin{figure}[htbp]
    \centering
  \includegraphics[width=150mm]{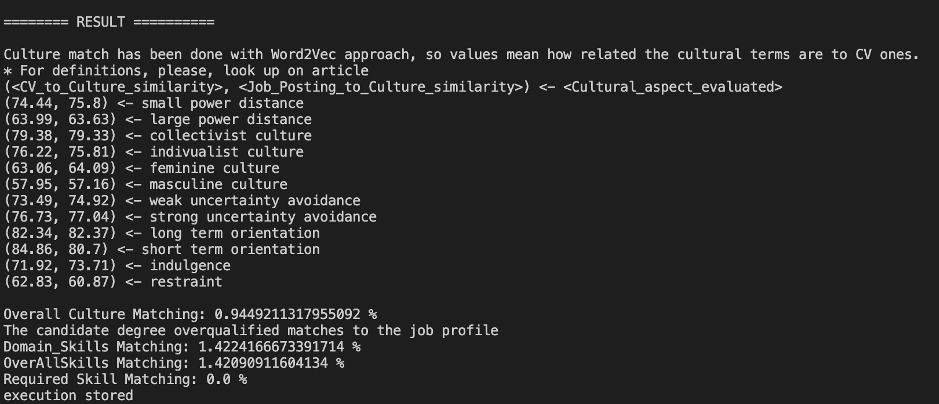}
  \caption{System output explaining the correlation between CV1 and a job post}
  \label{fig:result1}
\end{figure}

\subsubsection{Fig \ref{fig:result1} shows the result from test 3: CV1 and Business Oriented Job Post}

\begin{figure}[h!]
    \centering
  \includegraphics[width=150mm]{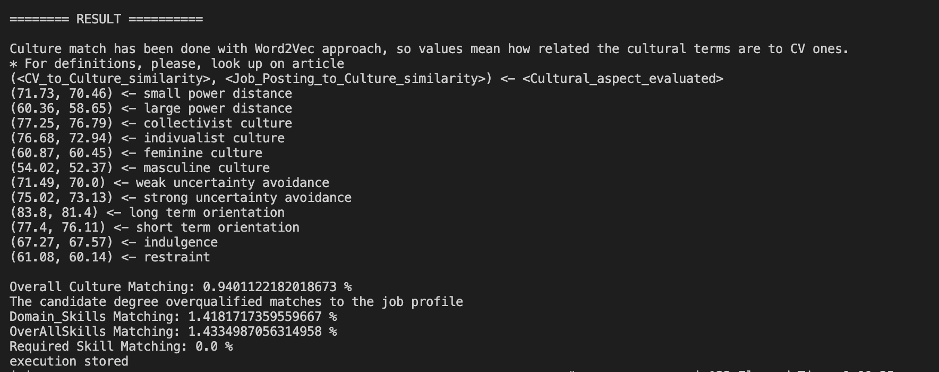}
  \caption{System output explaining the correlation between CV3 and a job post}
  \label{fig:result2}
\end{figure}

\subsubsection{Fig \ref{fig:result2} shows the result from test 4: CV3 and Technical Oriented Job Post}

\section{Conclusions}
About the process, a clear algorithm for recommendation could be implemented. Possible models for the CV parsing and recommendation processes are variate, not a complete approach can be found in research. We implemented a divide and conquer methodology for the model. We can approach each problem and solve each one with the best tools such as ontologies, embeddings, direct match, expert evaluation, machine learning. Develop an algorithm for the whole process according to the existence or not of data.
About the product, it would reduce time in the recruiting process, save money, invest recruiters in more productive activities in order to increase retention and productivity of the team and decrease recruiter’s bias. Besides, it would encourage best candidates' fitting on the organization, which would increase the company value as a consequence.

\section{Future Prospective}

The algorithm clearly shows a winning result as per the given time frame to complete the project. However there are many improvements which are possible in order to improve the result. In the scope of this project we explored mainly the domain of skill graph matching. The other important domain which is left unattended is the job titles and past experience. Adding the dimension of job title in current skill ontology can help to leverage deep match between candidates to the job positions. Moreover creating the word2vec model on the domain related data set will greatly help to explore closely the different concepts in the particular domain and to get the idea about their similarity and dissimilarity amongst them. In the scope of this project we did not consider the work experience of the candidate as we only focused to match candidates at a beginner level. However we wish to improve it further in order to adopt it for the experienced professional. Most importantly in order to deal with the cultural perspective of an organization we lacked the data from the organization as we believed that CV's of candidate accepted within the organization best describe the organization cultural values. By training the model on such data we could have achieved better results from a cultural perspective.

\section{Acknowledgment}
This work is supported by the researchers of IMT Atlantique, Brest. I would like to thank our professor Yanish Haralambous and Nicolas Julien who helped us for this study, and comments which helped to improve this paper.

% \section*{References}
\bibliographystyle{unsrt}
\bibliography{references}
\newpage
\section{Supplementary Information}

The pseudo code of the klink algorithm used to generate the CSO is below.

\begin{figure}[h!]
    \centering
  \includegraphics[width=100mm]{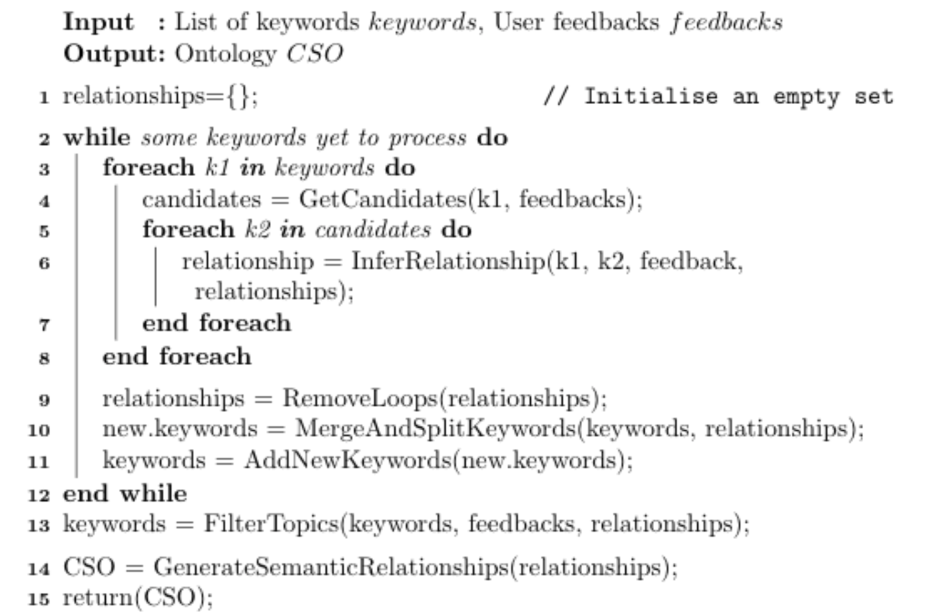}
  \caption{Data science doomain skill ontology }
\end{figure}

\begin{figure}[h!]
    \centering
  \includegraphics[width=150mm]{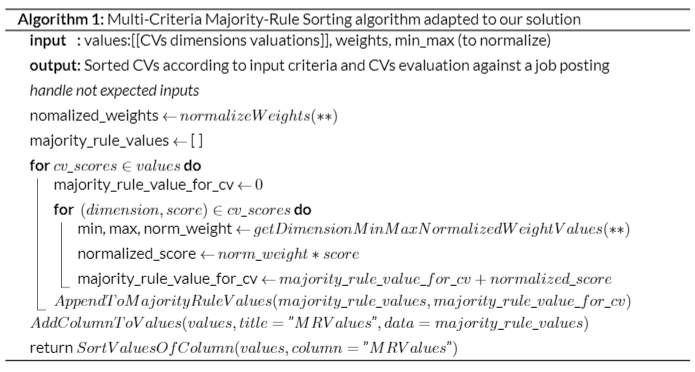}
  \caption{The pseudo code of the ML sort Algorithm }
\end{figure}

\end{document}